\documentclass{mn2e}

\usepackage[dvips]{graphicx}

\newcommand{\be}{\begin{equation}}
\newcommand{\ee}{\end{equation}}

\title[Wavelet analysis of MCG--6-30-15 and NGC 4051]
      {Wavelet analysis of MCG--6-30-15 and NGC 4051: a possible discovery of
       QPOs in 2:1 and 3:2 resonance}

\author
 [Lachowicz, Czerny \& Abramowicz]
{\noindent
 Pawe\l\ Lachowicz$^1$\thanks{E-mail: paulo@camk.edu.pl}, Bo\.zena Czerny$^1$ 
 and Marek A. Abramowicz$^{2,1}$
\\
$^1$Nicolaus Copernicus Astronomical Center, Bartycka 18, 00-716 Warszawa, Poland \\
$^2$Department of Physics, G\"{o}teborg University, S-412-96 G\"{o}teborg, Sweden\\
}

\pubyear{2006}

\begin{document}

\maketitle

\label{firstpage}


\begin{abstract}
Following our previous work of Lachowicz \& Czerny (2005), we explore further the 
application of the continuous wavelet transform to X-ray astronomical signals. Using
the public archive of the {\it XMM-Newton} satellite, we analyze all available EPIC-pn
observations for nearby Seyfert 1 galaxies MCG--6-30-15 and NGC 4051. We confine our analysis to
0.002--0.007 Hz frequency band in which, on the way of theoretically motivated premises,
some quasi-periodic oscillations (QPOs) are expected to be found. We find 
that indeed wavelet power histogram analysis reveals such QPOs centered at two 
frequencies of $\sim$$2.5\times 10^{-3}$~Hz and $\sim$$4\div 6\times 10^{-3}$~Hz, 
respectively. We show that these quasi-periodic features can be disentangled from the Poisson 
noise contamination level what is hardly to achieve with the standard Fourier analysis.
Interestingly, we find some of them to be in 2:1 or 3:2 ratio. If real, our finding may be 
considered as a link between QPOs observed in AGN and kHz QPOs seen in X-ray binary systems. 
\end{abstract}




\section{Introduction}
\label{s:s1}

One of the most remarkable features of the X-ray emission coming from active galactic 
nuclei (AGN) is their time variability. Rapid and aperiodic, rather than predictable
and deterministic (e.g. Lawrence et al. 1987; McHardy \& Czerny 1987; Mushotzky
et al. 1993), was quickly linked to processes taking place in the vicinity of the
central black-holes. In a recent widely accepted unified model of AGN 
(e.g. Urry \& Padovani 1995), soft X-rays/UV photons are emitted close to the central 
object in accretion disk and undergo upscattering via the inverse Compton process to 
higher energies in a hot plasma (e.g. Sunyaev \& Titarchuk 1980). 
Thus, information about intrinsic physical properties of the X-ray emitter is transmitted to
the observed X-ray flux which is affected by the Poisson noise process occurring as a 
result of their detection (e.g. van der Klis 1989).

The most classical way of description of broad-band variability of AGN is an 
estimation of the Fourier power spectral density (PSD). It gives an idea about the 
power of variable components (in terms of mean of the squared amplitude) as a 
function of frequency or, inversely speaking, at characteristic time-scales related
to the source. A commonly applied practice for all AGN is to study the 
overall shape of PSD by parameterizing it as a power law, 
$P(f)\propto f^{-\alpha}$, and 
quantifying it by the slope index
$\alpha$ which was found to be between 1 and 2 for most of Seyfert 1 galaxies (e.g.
Lawrence \& Papadakis 1993; Edelson \& Nandra 1999; Markowitz et al. 2003).

A weakness of the Fourier PSD approach to highly and aperiodically variable time-series
of AGN lies in its limited information that can be recovered. Since we 
are working only with a signal decomposition into frequency components, 
it is difficult to dig up an idea about e.g. their life-times and frequency distributions
that are responsible for the observed PSD shape. 
Vaughan et al. (2003) discussed fundamental tools for 
examination of AGN light curve properties along the time, however, they skipped  
the discussion on the methods pertaining to the studies of time-frequency domain.
So far, minor efforts have been undertaken in that direction for X-ray accretion astrophysics.
A promising tool to scan the local and shortly lasting quasi-periodic fluctuations is
associated with an application of the wavelet analysis (see e.g. Lachowicz \& Czerny 2005,
hereafter LC05, and references therein). It allows for a signal decomposition into 
two-dimensional space of parameters and to gather a knowledge about their deployment.

 \begin{figure}
 \vspace*{20pt}
 \includegraphics[width=8.2cm,angle=0]{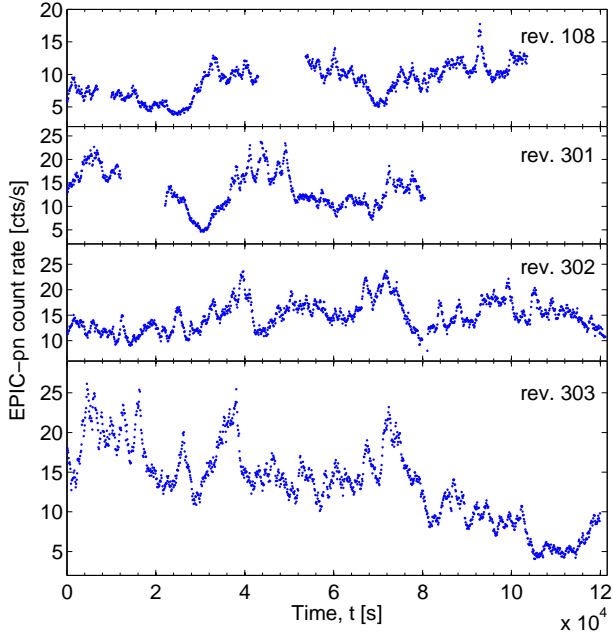}
 \caption
  {
   0.5--10 keV EPIC-pn light curves of MCG--6-30-15 analyzed in this paper with the wavelet 
   analysis. Here, for visual purposes, data were plotted with 100 s resolution.
  }
 \label{fig:lc}
 \end{figure}

Contrary to the results obtained for Galactic black-hole binary system of Cyg X-1 (LC05),
the usage of wavelet transform to X-ray data of AGN constitutes a challenging step towards 
understanding their time variability. As a subject of our study we choose MCG--6-30-15
and NGC 4051 which are frequently observed sources with a high count rate in X-ray energy band.

MCG--6-30-15 is a nearby Narrow-Line Seyfert 1 galaxy located at $z=0.007749$. It belongs to 
the one of the brightest in its own class. It shows up a very broad FeK line what might suggest 
that a central black-hole of a mass $2.9_{-1.6}^{+1.8} \times 10^6$ M$_\odot$ (McHardy et al. 
2005) is spinning rapidly and relativistic effects of 
space-time affect observed X-ray flux properties. A broad study of X-ray variability aspects for
MCG--6-30-15 have been done by Ponti et al. (2004) [rev. 108] and Vaughan, Fabian \& Nandra 
(2003) [rev. 301, 302, 303]. The latter found a typical PSD slope value of $\alpha\approx 2.5$ 
breaking to the flatter one ($\alpha\approx 1$) below $\sim 2\times 10^{-4}$ Hz. High-resolution 
PSD revealed a dominant white noise component at frequencies $\ga 5\times 10^{-3}$ Hz in 0.2--10 
keV band. NGC 4051 is another nearby Narrow-Line Seyfert 1 galaxy ($z=0.0023$) which displays a 
complex spectral behavior and highly variable X-ray flux (e.g. Ponti et al. 2006). Its PSD 
reveals energy independent bending at $8_{-3}^{+4}\times 10^{-4}$~Hz frequency changing a slope 
from -1.1 to about -2 (McHardy et al. 2004). Vaughan \& Uttley (2005) showed some evidences of 
QPO structure to be present at about $5\times 10^{-3}$~Hz which may be either real or arise due 
to inappropriate continuum model used in PSD fitting. A black-hole harbored in the heart of NGC 
4051 are considered to have a mass of $2\div 11\times 10^5$ M$_\odot$ (Shemmer et al. 2003; 
McHardy et al. 2004). 

 \begin{figure}
 \vspace*{20pt}
 \includegraphics[width=8.2cm,angle=0]{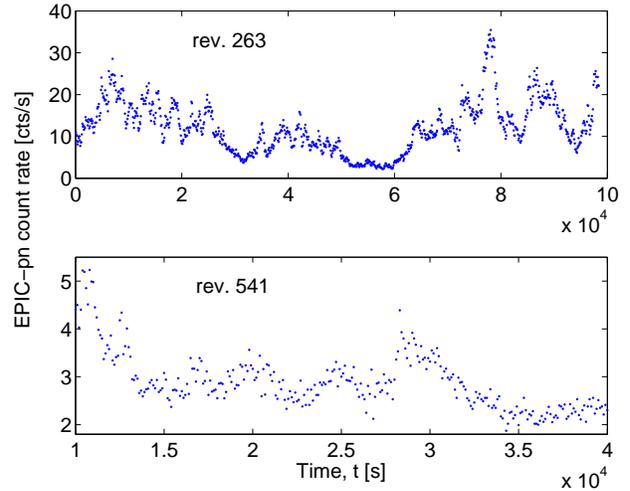}
 \caption
  {
   0.5--10 keV EPIC-pn light curves of NGC 4051 used in the wavelet analysis (plotted 
   with 100 s resolution).
  }
 \label{fig:lc2}
 \end{figure}

In this paper we apply the wavelet analysis to publicly available EPIC-pn/{\it XMM-Newton}
observations of MCG--6-30-15 and NGC 4051 in order to examine the time-frequency domain of
their variability. Our paper is organized as follows. In Section 2 we describe the details
of data selection and reduction. Section 3 contains a discussion of the wavelet analysis
issue stressing some crucial properties. The results of data analysis and their discussion
are presented in Section 4. We conclude in Section 5.


\section{Data selection and reduction}
\label{s:s2}

For almost all {\it XMM-Newton} (Jansen et al. 2001) observations of MCG--6-30-15
and NGC 4051, a pn 
(Strueder et al. 2001) and two MOS (Turner et al. 2001) CCD detectors of the European 
Photon Imaging Camera (EPIC) were operating simultaneously. Due to increased soft-$p^+$
flaring level detected in the data coming from the MOS cameras, we decided to ignore them
in our analysis. We processed all odf files with the {\it XMM-Newton} Science Analysis
Software (SAS v.7.0) using the most updated calibration files. Pn event lists were 
extracted using the standard processing SAS tool of {\sc epchain}. The source's counts were
read out from a circular region (centered for MCG--6-30-15 at 
${\rm R.A.}=13^{\rm h} 35^{\rm m} 53\fs 8$, $\delta=-34\degr 17' 44''$ and 
${\rm R.A.}=12^{\rm h} 3^{\rm m} 9\fs 6$, $\delta=+44\degr 31' 53''$ for NGC 4051)
of the radius equal $36\arcsec$ (rev.301) and $40\arcsec$ for the rest. A SAS tool {\sc 
tabgtigen} was applied to create good time interval (GTI) 
masks which were used in filtering the original pn event files. The background scientific 
products were extracted from the circular regions (of the same radius as used for the source)
on the same chip-set around the galaxies. The analysis of the pn data was performed using both
single and double events (i.e. pattern $\le 4$). In result, 0.5--10 keV light curve with 
time resolution of $\Delta T= 10$~s for each {\it XMM-Newton} data set was extracted and 
corrected for the background (Figure \ref{fig:lc} and \ref{fig:lc2}). We choose data analysis 
in that particular energy band to have highest possible signal-to-noise 
ratio\footnote{Energy-dependent wavelet analysis is the subject of our upcoming paper (Lachowicz 
et al., in preparation).}.
In addition, to work only with evenly distributed records, we applied an interpolation of that 
counts where X-ray flux: (a) dropped to zero, and/or (b) lay above $3\sigma$ around the mean 
(within a given segment; see below) and (c) a number of such nearby laid points was less then 
10. That correction comprised not more than 1 per cent of all data points in all seven data 
sets. Periods of distorted flux level (i.e. close to zero), continuous and longer than 200 s were 
skipped. 

For the purpose of the wavelet analysis, each resulting light curve\footnote{Hereafter,
we refer to each MCG--6-30-15 and NGC 4051 light curve by its {\it XMM-Newton} revolution 
number at which the galaxy was observed, e.g. rev. 301.} was divided into $T_{\rm seg}\simeq 
10^4$~s long segments what establishes a sample of data to be analyzed. Their total number, 
$M_{\rm seg}$, yielded 41 and 13 for MCG--6-30-15 and NGC 4051, respectively.

All performed calculations of the continuous wavelet transform were done using the {\sc 
matlab} software provided by Torrence \& Compo (1998, hereafter TC98) and Grinsted, 
Moore \& Jevrejeva (2004).


\section{Wavelet analysis}
\label{s:s3}

The wavelet analysis constitutes an alternative tool in a time-series analysis when 
confronted with the classical approach based on the Fourier techniques. In the latter,
a signal decomposition with the analyzing function, $\psi$, of the form of $e^{{\rm i}ft}$ 
allows to detect strictly periodic modulations buried in the signal. Signal integration 
in the Fourier transform is performed along its total duration. Thus, one cannot
claim anything about time-frequency properties of periodic components like their life-times
or time-localizations. A simple solution to this problem has been found via the application
of windowed Fourier transform (e.g. Cohen 1995), i.e. computation performed for a sliding 
segment of time-series (much shorter than a whole signal span) along the time axis. That 
returns a time-frequency chessboard of equal resolution everywhere dictated by the Heisenberg-Gabor 
uncertainty principle (Weyl 1928; Gabor 1946; Brillouin 1959; Flandrin 1999). A drawback of that 
method lies in the loss of information about fine and temporal quasi-periodic fluctuations 
(for example at high frequencies) which are simply smeared due to insufficient resolution. 

The wavelet analysis treats this problem more friendly as a basic concept standing for
its usefulness relay on probing time-frequency plane at different frequencies with
different resolutions. It can be achieved by selecting the analyzing function $\psi$
of finite duration in time that meets, in general, $\int_{-\infty}^{+\infty} \psi(t) dt=0$
condition. Making a scaling and time-shifting of $\psi$, at selected time duration of the
light curve, one can reach and examine a desired field of time-frequency space (see below).

Talking about wavelet analysis we mainly refer to the computation of a wavelet power spectrum
defined as the normalized square of the modulus of the wavelet transform:
\be
    W = \xi |w_k(a_m)|^2 .
\ee
Otherwise we stress the term we mean, for example, cross-wavelet spectrum or wavelet coherence 
(Kelly et al. 2003; Grinsted, Moore \& Jevrejeva 2004).

Here, for an introduction to the concept of continuous wavelet transform, we refer
a reader to our first paper of LC05 and references therein. We find there the best 
analyzing function (a mother wavelet) for X-ray signals to be the 
Morlet wavelet, $\psi(t)=\pi^{-1/4} e^{i 2\pi f_0 t} e^{-t^2/2}$, with $2\pi f_0$ parameter set 
to 6. It oscillates due to a term of $e^{{\rm i}t}$, has a complex form and
suits perfectly to probe a quasi-periodic modulations at different frequencies. 

Below, we extend our previous discussion on wavelet analysis (LC05) paying a special attention
to its usage in case of astronomical time-series following the Poisson distribution and to
general wavelet map characteristics.

\subsection{Heisenberg-Gabor uncertainty principle}
\label{sss:hg}

 \begin{figure}
 \vspace*{20pt}
 \includegraphics[width=8cm,angle=0]{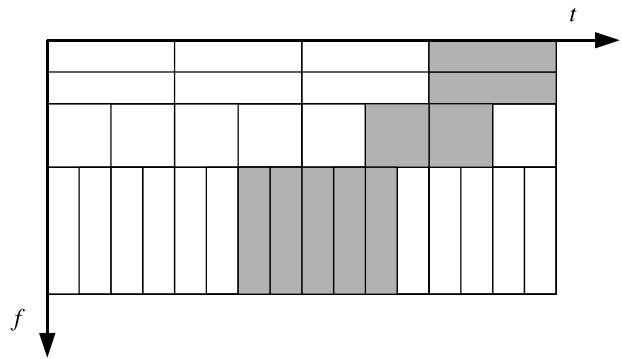}
 \caption
  {
   A chessboard of time-frequency signal decomposition in case of wavelet transform.
   Grey-color filled areas represent the exemplary non-zero coefficients of the
   wavelet decomposition with linearly increasing frequencies.
  }
 \label{f:board}
 \end{figure}

Despite a promising step forward, the wavelet analysis suffers from the same problem
encountered in the Fourier techniques, namely, the Heisenberg-Gabor uncertainty
principle. Applied to time-series it says that one cannot determine in an exact way
both time and frequency location of extracted features (wavelet peaks) simultaneously.
Certain uncertainty $\sigma$ exists and the product of frequency band-width $\Delta f$
with a time duration of the signal $\Delta t$ cannot be less than a minimal value given by:
\be
   \sigma^2 = \Delta t \Delta f \ge \frac{1}{4\pi}
\label{e:sigma2}
\ee
(e.g. Flandrin 1999). One can represent the spread of the wavelet in the time-frequency domain 
by drawing boxes of side lengths $2\Delta t$ by $2\Delta f$ known as Heisenberg boxes.
For the Gaussian windowed functions 
as used in the Morlet wavelet, the product of $\Delta t \Delta f$ is exactly equal $1/4\pi$.

An exemplary chessboard of time-frequency signal decomposition for the wavelet 
transform is shown in Figure \ref{f:board}. Shapes of the boxes represent the actual
resolution at the wavelet map in a function of linearly increasing frequencies. 
At high frequencies an increase in spread of spectral components is related to
decrease in a wavelet width in a time domain. A reverse situation takes place at lower
frequencies. Due to this fact, a sinusoidal fluctuation of the finite and the same time 
duration, located at frequencies $f_1, f_2, f_3$ where $f_1<f_2<f_3$, will differ in their
time-frequency properties in the wavelet map (Figure \ref{f:board}). We will return
to this point in Section \ref{sss:wpr}.

\subsection{Significance of wavelet peaks for Poisson signals}
\label{sss:poisson}

A map of wavelet power spectrum shows up a distribution of spectral components in the 
time-frequency plane. Single and finite oscillation, of sufficiently high amplitude to be 
detected, marks itself in the map as a (wavelet) peak. To every peak a certain statistical 
significance can be assessed. It was done for the first time by TC98. 
We say that a wavelet peak is significant with an assumed per cent confidence when it is
above a certain background spectrum. The latter is determined by the mean power spectrum of
analyzed time-series.

TC98 considered and adopted in the publicly available software the background power spectrum
of the following form:
\be
   P_j^{\rm AR1} = \frac{1-\alpha^2}{|1-\alpha e^{-i2\pi j}|^2}
\label{pj1}
\ee
which refers to an assumption that tested signal can be considered as a realization of
univariate lag-1 autoregressive (AR1) process:
\be
   x_l = \alpha x_{l-1} + z_l
\ee
where $\alpha$ is lag-1 autocorrelation, $z_l$ denotes a random variable drawn from the 
Gaussian distribution and $x_0=0$.
They showed that the distribution of the
local wavelet power spectrum follows $\chi^2_\nu$ distribution with $\nu$ degrees of freedom
(equal 1 or 2 for a real or complex wavelet, respectively).

However, a simple Lorentzian-shape power spectrum given by equation (\ref{pj1}) is not a good 
representation of a broad-band PSD shape of MCG--6-30-15 and NGC 4051 (e.g. Vaughan \& Uttley 
2005); the latters are better fitted with broken power-law models plus white noise.

In the present paper we aim at modeling very short light curves ($T_{\rm seg}\simeq 10^4 $ s, see
Section \ref{s:s2}) which are dominated by the Poisson white noise component.
Therefore, in order to define the background spectrum for a signal of Poisson distribution, 
we recall equation A7 of LC05 that defines, here, a normalized square of modulus of discrete 
form of continuous wavelet transform in terms of the Fourier transform as:
\be
   \xi |w_k(a_m)|^2\! =\! \left| \left(\frac{\xi2\pi a_m}{\Delta t} \!\right)^{1/2} 
   \sum_{j=1}^{N_{\rm 
   obs}}\!\hat{x}_j \hat{\psi}^\star(2\pi a_m\nu_j) e^{i2\pi\nu_j k\Delta t} \right|^2
\label{a:wkam}
\ee
where the discrete Fourier transform of signal $x_l$ is given by
\be
   \hat{x}_j = \sum_{l=0}^{N_{\rm obs}-1}
               (x_l - \bar{x}) e^{-i2\pi j l/N_{\rm obs}},
\ee
$j$ denotes a frequency index, $N_{\rm obs}$ a number of data points and $\bar{x}$
a mean value of $x_l$\footnote{We note that a factor $N_{\rm obs}^{-1}$ in 
equation A6 of LC05 appeared by mistake.}. Vaughan et al. (2003) presented a discussion
on popular normalizations of X-ray Fourier power spectra, namely $A|\hat{x}_j|^2$.
Among them we find $A=2\Delta t \bar{x}^{-1} N_{\rm obs}^{-1}$ as the most interesting 
for our purposes: (i) it has the property that for such normalized power spectrum
the expected Poisson noise level equals simply 2, and (ii) for any signal following the
Poisson distribution, resulting power spectrum should be distributed exactly as 
$\chi^2_2$, i.e. with 2 degrees of freedom (Leahy et al. 1983). That fact allows us thus
to construct the background spectrum for Poisson data as follows:
\be
   P_j^{\rm Poiss} = 2
\label{pj2}
\ee
assuming that the input light curve $x_l$ is in units of cts s$^{-1}$ and, what most
important, the normalization factor of the wavelet power spectrum equals:
\be
   \xi = AN_{\rm obs} = 2\Delta t \bar{x}^{-1} .
\label{xi}
\ee
Therefore the integrated wavelet spectrum over time (global wavelet spectrum)
defined as previously (LC05) in the form of:
\be
    G(a_m) = \frac{\xi}{N_{\rm obs}} \sum_k |w_k(a_m)|^2
\label{gam}
\ee
will hold the units of cts s$^{-1}$ Hz$^{-1}$, the same as $A|\hat{x}_j|^2$. Assuming 
normalization of (\ref{xi}) and recalling property (ii) given above, by the analogy to
TC98, the local wavelet power spectrum ought to be $\chi^2_2$ distributed, i.e.
\be
    \xi |w_k(a_m)|^2 \rightarrow \frac{1}{2} P_j^{\rm Poiss} \chi^2_2
\ee
for each time location $k$ and at particular scale $a_m$. In general, this is always
provided by the central limit theorem: the Fourier transform is the sum over time of 
a function of all $x_l$ and in the limit of large $N_{\rm obs}$ this sum tends asymptotically
to Gaussian distribution.

Summarizing the above, for shortly spanned X-ray light curve (provided widely e.g. by the 
detectors on board {\it XMM-Newton} or {\it RXTE} facilities) any wavelet peak standing 
significantly above the mean background spectrum given by (\ref{pj2}) can be considered as a 
true feature at assumed significance level\footnote{Fourier spectrum of the Poisson process
is frequency independent, thus $P_j^{\rm Poiss}$ does not depend on $j$. Here, we
neglect a red noise component since we focus at high frequencies.}.

 \begin{figure}
 \vspace*{20pt}
 \includegraphics[width=8cm,angle=0]{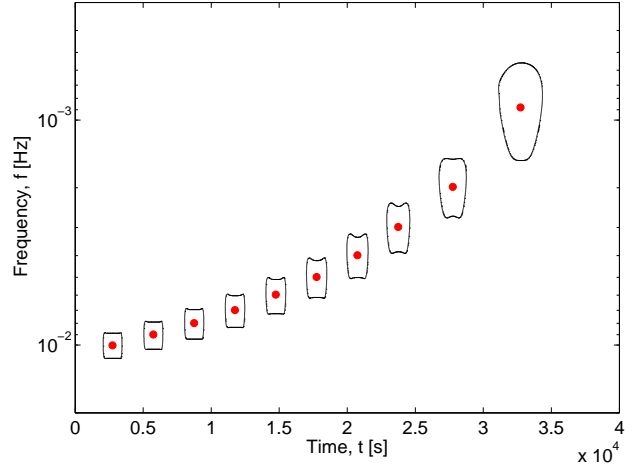}
 \caption
  {
    A simplified wavelet map for simulated 10 sinusoidal modulations of the finite time
    duration ($\delta t=1500$ s) localized at frequencies $f_i=0.001i$ Hz ($i=1,2,...,10$).
    Please note that frequency axis is plotted logarithmically contrary to Figure
    \ref{f:board}. See Section \ref{sss:wpr} for details.
  }
 \label{fig:sig}
 \end{figure}

\subsection{Statistical description of a wavelet map}
\label{sss:sta}

Depending on the frequency band that we wish to analyze with the wavelet analysis, 
a long-lasting light curve (of duration $T$ and time resolution $\Delta T$) is advised to
be divided into $M_{\rm seg}$ segments (of duration $T_{\rm seg}$).
Therefore, for every segment the wavelet power spectrum can be computed individually. In the 
resulting wavelet maps, as the wavelet peak one may define the contour of its statistical 
significance calculated at the given level. In order to describe quantitatively the content of 
peaks and their time-frequency properties along $T$, we propose:
\begin{enumerate}
\item
  to determine a median position of each peak: in time, $\lfloor t_0 \rceil$, and in 
  frequency, $\lfloor f_0 \rceil$;
\item
  to give the total number of peaks, $N_{\rm p}$, per light curve;
\item
  to determine the average number of peaks per segment, $N_{\rm p}/M_{\rm seg}$;
\item
  to read out a horizontal and vertical span and denote them as observed peak duration 
  $\delta t_{\rm obs}$ and its width in frequency $\delta f_{\rm obs}$, respectively, where 
  $\delta t_{\rm obs}$ should be corrected for the effect of the peak location at frequency
  $f$ (see Section \ref{sss:wpr} for details);
\item
  to calculate a median value of (sinusoidal) oscillations, $\lfloor N_{\rm osc} \rceil$,
  i.e. for case when assumed $\psi(t)$ is the Morlet wavelet and thus $N_{\rm osc}=\lfloor 
  f_0 \rceil\delta t$ for each peak.
\end{enumerate}
The above set of parameters can be treated as a basic one and used for the future purposes
of comparison of results from the wavelet analysis between different X-ray sources.

 \begin{figure}
 \vspace*{20pt}
 \includegraphics[width=8cm,angle=0]{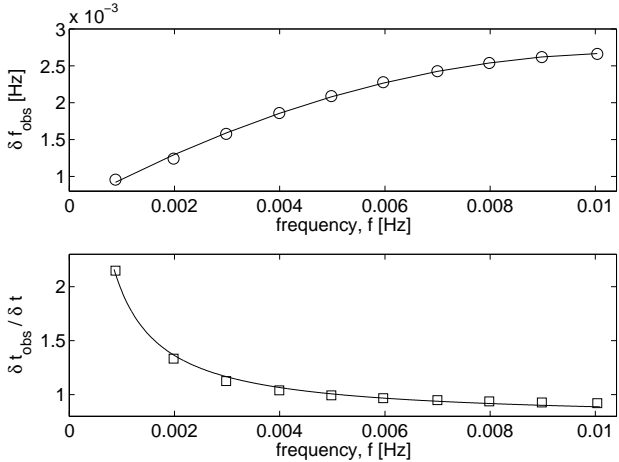}
 \caption
  {
    Upper: a dependence of frequency spread of the wavelet peak in a function of its
    location at frequency $f$ in the wavelet map. Bottom: a dependence of ratio of observed
    peak duration to the actual one in a function of signal localization at a given frequency
    (valid for $\Delta T=10$~s; see Section \ref{sss:wpr}).
  }
 \label{fig:dfftt}
 \end{figure}

\subsection{Wavelet peak resolution as a function of frequency}
\label{sss:wpr}

In order to quantify the dependencies of the signal duration $\delta t$ and its width in frequency 
domain, $\delta f$, as a function of the peak location at frequency $f$, the following 
simulation was performed. We generated a light curve of the total duration of 
$T=40000$ s at bin time $\Delta T=10$~s with 10 sine function $y_i=sin(2\pi f_i t)+const$, 
$f_i=0.001i$ and $i=1,2,...,10$. Each $y_i$ was assumed to last through a constant period of 
time of $\delta t=1500$ s, separated from each other by about 1500--2000 s in such a way that 
$const$ value was assumed between the sinusoidal parts of the signal. No noise component was 
added as we were interested only in a basic response of the wavelet transform to tested signals. 
A simplified wavelet map for that simulation is presented in Figure \ref{fig:sig}.

For a given wavelet peak its time-frequency location in the map was determined as given in
Section \ref{sss:sta} and marked by a dot in the figure. As tested before (LC05), due to 
irregularities in the peaks' shapes, in order to obtain best peak position, a median value 
was found to work better rather than a mean. Here, it returns a location in frequency with 
the average error of the order of $10^{-4}$, namely $f_i=0.001(\pm 0.0001)i$ for 
$i=2,3,...,10$.

In Figure \ref{fig:dfftt} (upper panel) the relation between $\delta f_{\rm obs}$ of the peak as 
a function of $f$ is given. A solid line represents the best fit to the data of the 
quadratic form:
\be
   \delta f_{\rm obs} = -18.11f^2 + 0.39f + 0.0006 .
\label{dfobs}
\ee
Therefore, a quantity of $\delta f_{\rm obs}/2$ can serve as a rough estimation of an error
we make in attempt of determination of the peak position at given frequency.

A second and non-negligible wavelet map property pertains to the dependence of the 
ratio of observed peak life-time to the actual one (i.e. assumed or/and known in advance)
as a function of $f$. It is not constant as one might expect. Figure \ref{fig:dfftt} 
(bottom panel) presents this relation.
Here, a solid line denotes the best fit of assumed model $y(f)= a/f + b$. Therefore,
one can correct the observed life-time of the wavelet peak, $\delta t_{\rm obs}$, 
for its real duration, $\delta t$, in the following way:
\be
   \delta t = \delta t_{\rm obs} \left( \frac{0.0012}{f} + 0.77 \right)^{-1} .
\label{dt}
\ee
Formula (\ref{dt}) can be treated as a good approximation of $\delta t$ with a relative
error about 2 per cent for the wavelet map calculation in 0.0008--0.01 Hz band. 

We would like also to note that relations (\ref{dfobs}) and (\ref{dt}) ought to be 
recalculated depending on our choice of $\Delta T$. Here, they are valid for $\Delta 
T=10$~s.


\section{Results and Discussion}
\label{s:results}

 \begin{figure}
 \vspace*{20pt}
 \includegraphics[width=8cm,angle=0]{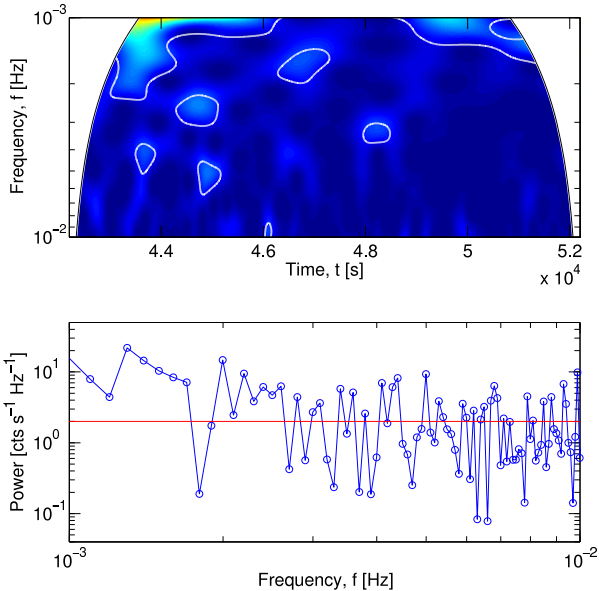}
 \caption
  {
    Upper panel: exemplary wavelet map calculated for $10^4$ s long segment of MCG--6-30-15 
    light curve (rev. 301). Top values of wavelet power are denoted by
    gradual brightening of the color. White solid contours denote significance level of
    98 per cent for extracted peaks. A cone of influence (COI) is marked as a white area at both
    sides of the map and corresponds to that space of wavelet spectrum for which the results are
    affected by the effect of signal finite duration. Bottom: corresponding Fourier power
    spectrum. Red straight line denotes the mean background spectrum at the expected Poisson 
    noise level.
  }
 \label{fig:wa4}
 \end{figure}

\begin{table*}
\caption{
   Properties of the localized peaks in the wavelet maps between 0.002--0.007 Hz
   (0.5--10 keV) for the assumed significance level of 98 per cent. $M_{\rm seg}$ denotes 
   the number of segments the given light curve was divided into. $N_{\rm p}$ stands for the 
   total number of detected wavelet peaks. $\lfloor\delta t \rceil$ and
   $\lfloor\delta f \rceil$ are the median values of wavelet peak life-times and frequency 
   widths, respectively, whereas $\lfloor N_{\rm osc} \rceil$ is the median value of the 
   sinusoidal oscillations among all peaks. $\bar{x}$ denotes the mean count rate of the
   light curve.
}
\label{tab1}
\begin{tabular}{crrrcccc}
\hline
Rev. &
$M_{\rm seg}$ &
$N_{\rm p}$ &
$N_{\rm p}/M_{\rm seg}$ &
$\lfloor\delta t \rceil$ &
$\lfloor\delta f \rceil$ &
$\lfloor N_{\rm osc} \rceil$ &
$\bar{x}$ \\
&&&&(s)&(Hz)&&(cts s$^{-1}$)\\
\hline
\multicolumn{8}{c}{\it MCG--6-30-15} \\
   108 &  9 &  26 & 2.89 & 376.3 &  0.0017 &  2.04 & 8.9  \\
   301 &  8 &  24 & 3.00 & 499.6 &  0.0011 &  1.89 & 13.5 \\
   302 & 12 &  40 & 3.33 & 405.9 &  0.0013 &  1.75 & 14.9 \\
   303 & 12 &  39 & 3.25 & 286.7 &  0.0009 &  1.11 & 13.6 \\
\multicolumn{8}{c}{\it NGC 4051} \\
   263 & 10 &  76 & 7.60 & 354.8 &  0.0013 &  1.33 & 12.1 \\
   541 &  3 &   8 & 2.67 & 703.7 &  0.0019 &  2.34 &  2.9 \\
\hline
\end{tabular}
\end{table*}

We construct the wavelet power spectrum maps for $M_{\rm seg}$ segments of data and characterize
the source activity during a given period by extracted wavelet peaks at the significance level 
of 98 per cent\footnote{We explain our choice of that particular significance level in 
Section~\ref{ss:statt}.}. Figure \ref{fig:wa4} (upper panel) presents an exemplary outlook of such 
time-frequency decomposition performed for one segment of data (rev. 301). In the bottom panel 
of the same figure we show the corresponding Fourier power spectrum calculated for the same 
frequency band as given in the wavelet map. The red straight line represents the mean background 
spectrum expected to be the Poisson noise level (Section \ref{sss:poisson}). For every light 
curve we calculate its statistical characteristics as described in Section \ref{sss:sta}
and summarize them in Table \ref{tab1}. In Table \ref{tab2} we compare globally the properties
between MCG--6-30-15, NGC 4051 and simulated light curves of Poisson noise.

\subsection{Statistical properties of wavelet activity}
\label{ss:statt}

Table \ref{tab2} reveals that the median life-times of peaks for each object remain
very similar, however, they change within short period of time (Table \ref{tab1}). In 
particular, for MCG--6-30-15 observed 3 times during consecutive 5 days (rev. 301--303), the 
maximal difference in $\lfloor\delta t \rceil$ yields about 213 s. More dramatic changes we can 
see for NGC 4051, i.e. of about 350~s, but, in this case, separated by 18 months in the source 
observation moments. That may indicate that character of physical processes operating in the 
source at time-scales between 142~s and 500~s is not constant and may change even by a factor of 
$\sim$2 or more for $\lfloor\delta t\rceil$ within a few days. 

\begin{figure}
 \vspace*{20pt}
 \includegraphics[width=8cm,angle=0]{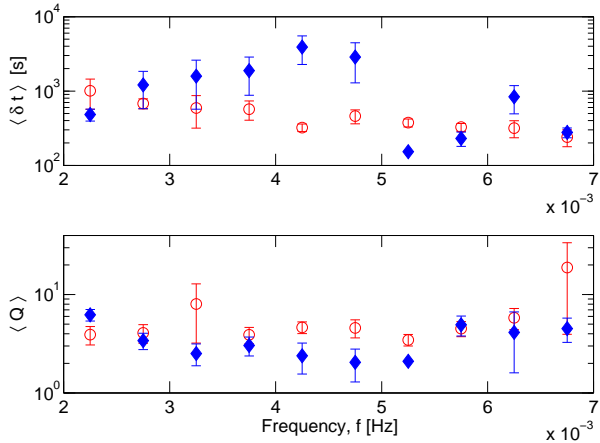}
 \caption
  {
    A relation between (upper panel) the average life-time of wavelet peaks, $\langle \delta t
    \rangle$, and (bottom part) the average quality factor, $\langle Q \rangle$, of detected
    wavelet peaks as a function of frequency for MCG--6-30-15 (open circles) and NGC 4051
    (filled diamonds).
  }
 \label{fig:lifetime}
 \end{figure}

Interestingly, the average life-time of the peaks does not seem to be constant function
of frequency at which they occur. It is shown in the upper panel of Figure \ref{fig:lifetime}.
For MCG--6-30-15 the existing variability structures tend to live longer at longer
time-scales but for NGC 4051 the same behavior is different. Oscillations live longer with
increasing frequency, peaking at about $4.5\times 10^{-3}$~Hz and then suddenly drop down
by a factor of 20. The cause of this rapid drop is unclear at this stage of research.

The same cannot be read out for the ratio of the peak frequency to its width, denoted widely
in the literature as a quality factor, $Q=f/\delta f$. For both sources is remains 
approximately constant with the average value of $\langle Q\rangle\simeq 3$.

\begin{figure}
 \vspace*{20pt}
 \includegraphics[width=8cm,angle=0]{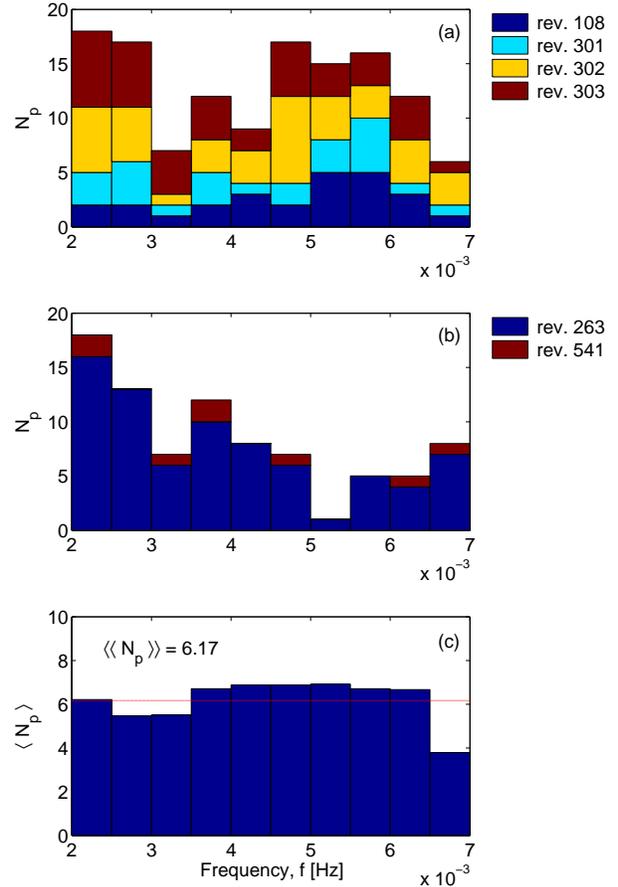}
 \caption
  {
    Histogram of wavelet peak distribution in a function of their location
    in frequency: (a) for MCG--6-30-15 where colors refer to the single histogram
    values for each light curve separately (see legend) whereas joined together
    represent a contribution to the total peak activity in a given frequency band,
    (b) for NGC 4051, and (c) for simulated time-series of the pure Poisson noise
    (see Section \ref{ss:pure} for details).
  }
 \label{fig:barplot}
 \end{figure}

The results obtained from the wavelet analysis also provides us with the opportunity of the 
histogram representation (hereafter also referred to as the wavelet power histogram)
of the total number of detected oscillations at a given frequency. It 
can be considered as an analogue to the global wavelet power spectrum or the Fourier power 
spectrum. We performed a computation of such histograms for every light curve separately and
displayed the result in the form of stacked bar graph for MCG--6-30-15 (Figure 
\ref{fig:barplot}a) and NGC 4051 (Figure \ref{fig:barplot}b), respectively.
These graphs show up contributing amounts of the wavelet peaks detected in each observation to 
their overall activity at given frequency. As denoted by Ponti et al. (2006), the second {\it 
XMM-Newton} observation of NGC 4051 (rev. 541) was timed to coincide with an extended period of 
its low X-ray emission that is why the total number of extracted events from this 
observation is very low (Table \ref{tab1}).

\begin{table}
\caption{
   Comparison of the wavelet peak characteristics obtained from the analysis of
   MCG--6-30-15 and NGC 4051 data and simulated time-series containing only a pure Poisson
   noise. Rows
   (up to down) provide information on: number of segments, total number of extracted
   wavelet peaks, the average number of peaks per segment, the median life-time of peaks,
   the median width of peaks and the median number of sinusoidal oscillations within
   duration of the peak, respectively. See Section \ref{ss:pure} for details.
}
\label{tab2}
\begin{tabular}{llll}
\hline
Quantity &
MCG--6-30-15 &
NGC 4051 &
Simulation
\\
\hline
$M_{\rm seg}$                  & 41        & 13         & 1000  \\
$N_{\rm p}$                    & 129       & 84         & 1553          \\
$N_{\rm p}/M_{\rm seg}$        & 3.14      & 6.46       & 1.55     \\
$\lfloor\delta t \rceil$       & 366.6 s   & 385.6 s    & 308.5 s   \\
$\lfloor\delta f \rceil$       & 0.0012 Hz & 0.0012 Hz  & 0.0011 Hz\\
$\lfloor N_{\rm osc} \rceil$   & 1.56      & 1.48       & 1.33      \\
\hline
\end{tabular}
\end{table}

For MCG--6-30-15 the histogram reveals a sort of bimodal distribution with a distinctive peak at
$\sim$$2.5\times 10^{-3}$~Hz and a broad one centered at $\sim$$5\times 10^{-3}$~Hz,
respectively. It is difficult to judge its reality looking only at the
Fourier power spectra (cf. figure 4 in Vaughan, Fabian \& Nandra 2003). However, it is worth
to mark that our high frequency (HF) peak stays in a good agreement with some predictions for 
HF QPO (to be in $0.5-3\times 10^{-3}$~Hz) of Vaughan \& Uttley (2005). A less 
pronounced bimodal character can be seen for the NGC 4051 wavelet power histogram. Here, an 
indication of QPO at $5\times 10^{-3}$~Hz in NGC 4051 (mentioned in Section \ref{s:s1}) differs 
comparing to our results. 

Histograms for both galaxies were calculated at assumed significance level of 98 per cent. 
We have checked the results for 95--99 per cent band of significance and
we have found that only for 99 per cent level the histograms failed to show up a bimodal
distribution due to lowest number of extracted peaks. Therefore, 98 per cent level was
assumed by us as the highest possible level in order to perform calculations and the
analysis of results.

 \begin{figure}
 \vspace*{20pt}
 \includegraphics[width=8cm,angle=0]{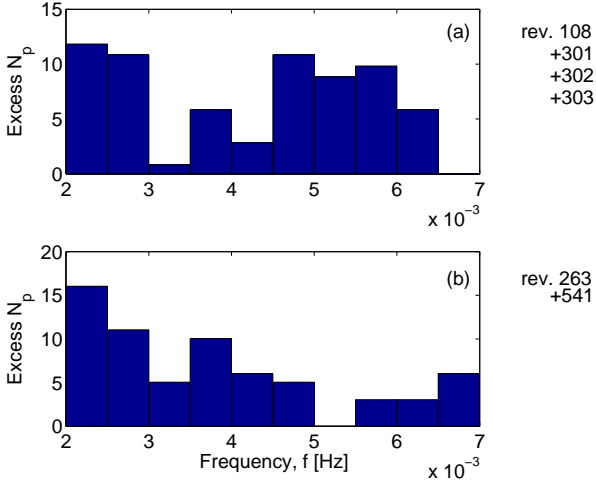}
 \caption
  {
    Excess wavelet power histogram for combined (a) MCG--6-30-15 and (b) NGC 4051 
    light curves, respectively. See Section \ref{ss:pure} for details.
  }
 \label{fig:barplot2}
 \end{figure}

\subsection{Wavelet analysis of pure Poisson noise and an excess wavelet power histogram}
\label{ss:pure}

Vaughan, Fabian \& Nandra (2003) performed a detailed Fourier power spectrum analysis
for MCG--6-30-15 observations at rev. 301, 302 and 303. For high-frequency periodograms
estimated using $\Delta T=10$~s they found that a red-noise PSD of the source can be
seen for frequencies $\la 3\times 10^{-3}$~Hz. It is intuitively confirmed in the bottom
part of Figure \ref{fig:wa4}, though, in this case $T_{\rm seg}\simeq T/8$, i.e. time-series
is much shorter than a whole light curve duration. Therefore, based on the Fourier technique
results one can consider 0.003--0.007 Hz fully dominated by the Poisson noise. Hereby, we 
challenge this statement to may be not fully correct by performing the following
Monte-Carlo simulation.

To find out about the ratio of all wavelet peaks appearing in the wavelet maps due to pure
Poisson noise to the total number of peaks observed for MCG--6-30-15 (and NGC 4051), 
we generated 1000 light curves of duration $10^4$~s each at $\Delta T=10$~s. Here, we 
followed the same procedure as described in section 5.1 of LC05 (equation 13), however assuming
a constant value equal to the average count rate in MCG--6-30-15 and NGC 4051 EPIC-pn data, i.e. 
$\simeq$15 cts~$s^{-1}$ (0.5--10 keV). For every single time-series the wavelet power spectrum in 
0.002--0.007~Hz frequency band was calculated and statistical characteristics (of Section 
\ref{sss:sta}) were done.

Table \ref{tab2} confronts the wavelet peak properties between these obtained for two
galaxies (all light curves included) and from the simulation. The most striking result
pertains to the ratio of the average number of peaks (per segment) extracted for 
MCG--6-30-15 and for pure Poisson process realization\footnote{All detected wavelet peaks 
for Poisson noise signals correspond to real and significant wavelet powers
standing out above the {\it mean} background spectrum.} which is equal 
to $\sim$2. Therefore, this ratio says that statistically about half of all observed wavelet 
peaks in 0.5--10 keV X-ray MCG--6-30-15 emission is of the pure noise origin. NGC 4051
occurs to be more active with the average number of peaks per segment of $\sim$6.5 where
only 24 per cent of them can arise due to noise.

In order to visualize the distribution of wavelet peaks being a pure noise contamination,
we construct the noise histograms in the following way. We group $m\times M_{\rm seg}$
simulated light curves of Poisson noise into $m$ parts. For each part (composed of 
$M_{\rm seg}$ segments) first we sort the list of peaks according to frequency and then
we calculate a histogram (10 bins with 0.0005 Hz bin-width). Finally, for all of $m$ histograms, 
in a given histogram-bin we compute the mean number of peaks, $\langle N_p \rangle$, and 
associated error. The latter appears to be of the order of $10^{-2}$.
Here, $m$ was assumed to be equal 24 and 76 for MCG--6-30-15 and NGC 4051, 
respectively. In result we found that both histograms revealed a uniform distribution of 
peaks as a function of frequency with the mean value, $\langle\!\langle N_p \rangle\!\rangle=
6.17$ and 1.96 for $m=24$ and 76, respectively. Figure \ref{fig:barplot}c shows 
up the noise histogram corresponding to $m=24$ and $M_{\rm seg}=41$ (the error bars have been
omitted).

This outcome can serve us as a good estimation of the average number of wavelet peaks in the 
light-curves that are of a pure noise origin. Therefore, it also allows us to subtract 
$\langle\!\langle N_p \rangle\!\rangle$ value from the data (Figure \ref{fig:barplot}a,b) 
obtaining an {\it excess wavelet power histogram} presented in Figure \ref{fig:barplot2} for 
both galaxies. Thus, the excess histogram provides us with information on the wavelet peak 
distribution in the source disentangled from the average level of wavelet peaks arising
only due to noise process.

 \begin{figure}
 \vspace*{20pt}
 \includegraphics[width=8cm,angle=0]{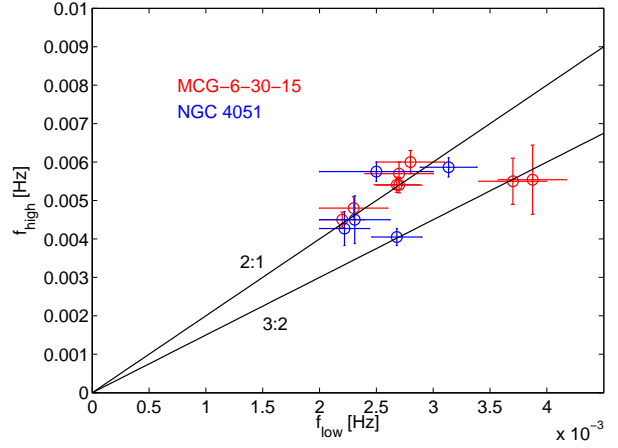}
 \caption
  {
   Frequency-frequency plots for twin peak QPOs in MCG--6-30-15 and NGC 4051.
   The plots are remarkably similar. They suggest that two resonances may be present here,
   the 3:2 resonance that is also common in the Galactic black hole and neutron star sources, 
   and a 2:1 resonance. 
  }
 \label{fig:resonance}
 \end{figure}

\subsection{The frequency-frequency ($f_{\rm high}$ vs. $f_{\rm low}$) plot}
\label{ss:ffplot}

Quasi-bimodal distribution of peaks that can be seen in Figure \ref{fig:barplot2}a
suggests that MCG--6-30-15 may show double peak QPOs (with frequencies $f_{\rm high}$ and 
$f_{\rm low}$) as indeed many Galactic black hole and neutron sources do. One can read out from 
this figure that the ratio is equal about $f_{\rm high}:f_{\rm low}\simeq 2:1$ assuming $f_{\rm 
high}$ and $f_{\rm low}$ to be $\sim$0.0055~Hz and $\sim$0.0025~Hz, respectively. If the 
histogram peak at $\sim$0.00375~Hz goes into a play that  indicates also at 3:2 frequency ratio. 
Similar ratios can be found for the excess histogram of NGC 4051 (Figure \ref{fig:barplot2}b).

We decided to investigate this case more carefully. We divided the list of wavelet peaks
detected in each of four MCG--6-30-15 light curves into two parts of equal amounts.
Thus, for 8 parts of data we calculated 8 histograms (10 bins) of the peak distribution as a
function of frequency. Each histogram revealed two distinctive and well separated peaks
at frequencies denoted here as $f_{\rm low}$ and $f_{\rm high}$, respectively. In order
to increase the precision of frequency determination we also checked the histograms with 
11 and 12 bin resolution. Where needed, 8 and 9 bin widths were also investigated. By the average 
error of determination of peak position we assumed a half of the peak width. In result, we 
plotted the relation of $f_{\rm high}$ as a function of $f_{\rm low}$ for given 8 pairs in one 
figure (see Figure \ref{fig:resonance}). We have repeated the same procedure for NGC 4051
(5 pairs) obtaining very similar results.

Rather remarkably, for both sources a few points locate along a 3:2 frequency ratio line, 
and the rest close to a 2:1 line. No other combinations are observed. 

\subsection{Double peak QPOs in MCG--6-30-15 and NGC 4051: a link to Galactic kHz~QPOs?}
\label{ss:link}

 \begin{figure}
 \vspace*{20pt}
 \includegraphics[width=8cm,angle=0]{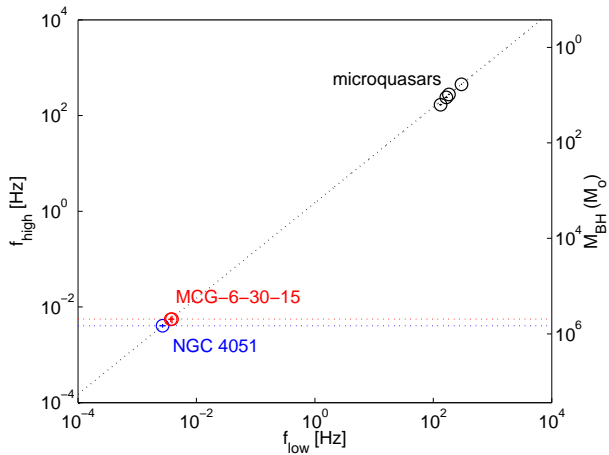}
 \caption
  {
   McClintock \& Remillard (2003) scaling for microquasars extended
   down to AGN region with marked results found for MCG--6-30-15 and NGC 4051 within
   our wavelet research. If real, it may be considered as the link between QPOs observed in 
   AGNs and kHz QPOs seen in X-ray binary systems (see Section \ref{ss:link} for discussion).
  }
 \label{fig:mscaling}
 \end{figure}

Results discussed in the previous Section and displayed in Figure \ref{fig:resonance} suggest
that double peak QPOs in both MCG--6-30-15 and NGC 4051 have {\it commensurable} frequency
ratios, 3:2 and 2:1. Abramowicz \& Klu{\'z}niak (2001) first noticed the 3:2 ratio (and stressed
its importance) in the X-ray variability data of the microquasar GRO J1655-40 analyzed by
Strohmayer (2001); see also McClintock \& Remillard (2003). Abramowicz et al. (2003) argued that
the 3:2 ratio is also evident in the case of the neutron star QPOs. Although the ratio is
varying, it clearly distinguishes the 3:2 value (see Belloni et al. 2005 for a
criticism). There is a consensus today that the commensurable frequency ratios, and 3:2 in
particular, are an important property of the QPO phenomenon in microquasars and neutron star
sources (for recent references see T{\"o}r{\"o}k et al. 2005). It has been pointed out by
Klu\'zniak \& Abramowicz (2000) that a commensurable frequency ratio suggests a resonance, 
and that the double peak QPO in microquasars and neutron stars are due to a {\it similar} 
non-linear resonance between two modes of accretion disk oscillations in strong gravity.

Our results may suggest that the same 3:2 resonance operates also in the MCG--6-30-15 and
NGC 4051 accretion discs. The full theoretical discussion of this suggestion will be 
presented in a follow-up paper, now in preparation. Here, we would like to make only one 
comment, on the QPO frequency scaling with the inverse mass of the source, that is 
theoretically expected in strong gravity. McClintock \& Remillard (2001) noticed that 
for three microquasars with known mass (see also T{\"o}r{\"o}k et al. 2005), the QPO 
frequency indeed scales according to:
\be
f_{\rm high} = 2793 \left( \frac {M}{M_{\odot}} \right)^{-1} \,\rm{Hz}.
\label{1/M}
\ee

In Figure \ref{fig:mscaling} we show that this scaling holds for NGC 4051 with the 
commonly adopted mass $M = 2\div 11\times 10^5$~M$_{\odot}$ (Shemmer et al. 2003; McHardy et 
al. 2004), but for MCG--6-30-15 with a rather low mass estimate, $M\simeq 5\times 
10^5$~M$_{\odot}$ that according to McHardy et al. (2005) is possible, but not very likely.


\section{Conclusions}
\label{s:conclusions}

We have presented the results of the wavelet analysis for two Narrow-Line Sy1 galaxies, 
MCG--6-30-15 and NGC 4051, in 0.5--10 keV energy band. The main points of our research can be 
summarized as follows:
\begin{enumerate}
\item
     We explored the observational properties of time-frequency wavelet chessboard
     (Section \ref{s:s3}) and proposed a set of common parameters that can be used in 
     the future in order to compare wavelet power spectrum results obtained for different 
     X-ray sources (Section \ref{sss:sta});
\item
     We discussed and showed (Section \ref{sss:poisson}) that the most proper mean background 
     spectrum required in the tests for the significance of wavelet peaks in case of Poisson 
     signals can be assumed as given by equation (\ref{pj2});
\item
     We showed that the wavelet (histogram) analysis is a powerful and efficient method to 
     extract real signals buried in the Poisson noise which are present in the data for very 
     limited fraction of time, and how to disentangle them from that noise (Section 
     \ref{ss:pure});
\item
     Finally, we seem to detect pairs of QPO oscillations in MCG-6-30-15 and NGC 4051, which
     are either in 2:1 or 3:2 resonance (Section \ref{ss:ffplot}). If confirmed in the future 
     with extended number of observations for the same and/or other bright AGNs, our finding may 
     indicate on a link between QPOs observed in AGNs and kHz QPOs seen in X-ray binary systems
     (Section \ref{ss:link}).
\end{enumerate}


\section*{ACKNOWLEDGMENTS}

PL thanks Simon Vaughan for friendly and valuable discussions on the issue of power spectra 
of an AR1 process and Aslak Grinsted on the wavelets. PL would like also to thank 
David Guetta for his fantastic song {\it The world is mine} which was a great motivation and 
inspiration during the work over this project. Part of this work was supported by grant 
1P03D00829 of the Polish State Committee for Scientific Research. The wavelet software was 
provided by Christopher Torrence and Gilbert P. Compo and it is available at 
http://paos.colorado.edu/research/wavelets.


\bsp
\label{lastpage}

\begin{thebibliography}{}

\bibitem{AK01}
Abramowicz M.A., Klu\'zniak W., 2001, A\&A, 374, L19

\bibitem{abr03}
Abramowicz M.A., Bulik T., Bursa M., Klu\'zniak W., 2003, A\&A, 404, L21

\bibitem[2005]{bel05}
Belloni T., M{\'e}ndez M., Homan J., 2005, A\&A, 437, 209

\bibitem{brill59}
Brillouin L., 1959, La Science et la Theorie de l'Information, Paris: Gauthiers-Villars
 
\bibitem{cohen95}
Cohen L., 1995, Time-frequency analysis, Englewood Cliffs, NJ: Prentice-Hall

\bibitem{ede99}
Edelson R., Nandra K.,  1999, ApJ, 514, 682

\bibitem{flandrin99}
Flandrin P., 1999, Time-Frequency/Time-Scale Analysis, Academic Press

\bibitem{gabor46}
Gabor D., 1946, Theory of communication, J.IEE, 93, 429

\bibitem{gilma63}
Gilman D.L., Fuglister F.J., Mitchell J.M. Jr.,  1963, JAtS, 20, 182

\bibitem{gri04}
Grinsted A., Moore J.C., Jevrejeva S.,  2004, NPGeo, 11, 561

\bibitem{janlu}
Jansen, F., Lumb, D., Altieri, B., Clavel, J., Ehle, M., et al., 2001,
 A\&A, 365, L1

\bibitem{kelly03}
Kelly B.C., Hughes P.A., Aller H.D., Aller M.F., 2003, ApJ, 591, 695

\bibitem{klu00}
Klu\'zniak W., Abramowicz M.A., 2000, Phys. Rev. Lett. (submitted), astro-ph/0105057

\bibitem{lc05}
Lachowicz P., Czerny B., 2005, MNRAS, 361, 645 (LC05)

\bibitem{law87}
Lawrence A., Watson M.G., Pounds K.A., Elvis M., 1987, Nature 325, 694

\bibitem{law93}
Lawrence A., Papadakis I., 1993, ApJ, 414, L85

\bibitem{mark03}
Markowitz A., Edelson R., Vaughan S., et al.,  2003, ApJ, 593, 96

\bibitem{mch87}
McHardy I., Czerny B., 1987, Nature 325, 696

\bibitem{mch04}
McHardy I.M., Papadakis I.E., Uttley P., Page M.J., Mason K.O., 2004, MNRAS, 348, 783

\bibitem{mch05}
McHardy I., Gunn K.F., Uttley P., Goad M.R.,  2005, MNRAS, 359, 1469

\bibitem{mccrem}
McClintock J.E., Remillard R.A., 2004, in Compact Stellar X-ray Sources, ed. W. H. G. Lewin \& M. 
van der Klis (Cambridge: Cam. Univ. Press), astro-ph/0306213

\bibitem{mush93}
Mushotzky R.F., Done C., Pounds K.A., 1993, ARA\&A 31, 717

\bibitem{ponti04}
Ponti G., Cappi M., Dadina M., Malaguti G.,  2004, A\&A, 417, 451

\bibitem{ponti06}
Ponti G., Miniutti G., Cappi M., Maraschi L., Fabian A.C., Iwasawa K., 2006, 
MNRAS, 368, 903

\bibitem{shemm03}
Shemmer O., Uttley P., Netzer H., McHardy I.M., 2003, MNRAS, 343, 1341

\bibitem{sun80}
Sunyaev R.A., Titarchuk L.G., 1980, A\&A 86, 121

\bibitem{stro01}
Strohmayer T., 2001, ApJ, 554, L169

\bibitem{strudel}
Strueder, L., Briel, U., Dennerl, K., Hartmann, R., Kendziorra, E.,
 2001, A\&A, 365, L18

\bibitem{torok05}
T{\"o}r{\"o}k G., Abramowicz M.A., Klu\'zºnia, W,; StuchiÃk, , 2005, A\&A, 436, 1

\bibitem{tc98}
Torrence C., Compo G.P., 1998, BAMS, 79, 61

\bibitem{turabb}
Turner, M.J.L., Abbey, A., Arnaud, M., Balasini, M., Barbera, M., et al.,  2001,
A\&A, 365, L27

\bibitem{urry95}
Urry C.M., Padovani P., 1995, PASP, 107, 803

\bibitem{vdk89}
van der Klis M., 1989, ARA\&A, 27, 517

\bibitem{vau03a}
Vaughan S., Edelson R., Warwick R.S., Uttley P.,  2003, MNRAS, 345, 1271

\bibitem{vfn03}
Vaughan S., Fabian A.C., Nandra K.,  2003, MNRAS, 339, 1237

\bibitem{vautt05}
Vaughan S., Uttley P., 2005, MNRAS, 362, 235

\bibitem{weyl28}
Weyl H., 1928, Gruppentheorie un Quantenmechanik, Leipzig: S. Hirzel

\end{thebibliography}
\end{document}